\documentclass[a4paper,11pt]{article}

\usepackage[usenames,dvipsnames]{color}
\usepackage{amssymb,amsmath,eepic,graphics,color,enumerate}
\usepackage{tikz}
\usetikzlibrary{calc}

\usepackage{epsfig}

\usepackage[T1]{fontenc}

\newcommand{\proper}{\ensuremath{\mathrm{proper}}}

\usepackage{amsmath,amsthm,amsfonts,amssymb,epsfig,color,xy}

\newtheorem{theorem}{Theorem}[]

\newtheorem{corollary}[theorem]{Corollary}
\newtheorem{proposition}[theorem]{Proposition}

\renewcommand{\subset}{\subseteq}

\begin{document}
\title{Channel Assignment via Fast Zeta Transform}

\author{Marek Cygan and \L{}ukasz Kowalik\\
Institute of Informatics, University of Warsaw \\
{\tt \{cygan,kowalik\}@mimuw.edu.pl}
}
\date{}

\maketitle

{%\linespread{1.2}
%\maketitle

%\medskip\noindent{\bf Keywords:} 3-edge-coloring, subgraph, approximation

\newcommand{\heading}[1]{\medskip\noindent\textbf{#1}.\ } 
\newcommand{\headingnsp}[1]{\noindent\textbf{#1}.\ } 

\begin{abstract}
We show an $O^*((\ell+1)^n)$-time algorithm for the channel assignment problem, where $\ell$ is the maximum edge weight.
This improves on the previous $O^*((\ell+2)^n)$-time algorithm by Kral~\cite{kral},
as well as algorithms for important special cases, like $L(2,1)$-labelling. 
For the latter problem, our algorithm works in $O^*(3^n)$ time.
The progress is achieved by applying the fast zeta transform 
in combination with the inclusion-exclusion principle.
\end{abstract}}

\section{Introduction}
In the channel assignment problem, we are given a symmetric weight function $w:V^2\rightarrow \mathbb{N}$ (we assume that $0\in\mathbb{N}$).
The elements of $V$ will be called vertices (as $w$ induces a graph on the vertex set $V$ with edges corresponding to positive values of $w$).
We say that $w$ is $\ell$-bounded when for every $x,y\in V$ we have $w(x,y)\le \ell$. An assignment $c:V\rightarrow\{1,\ldots,s\}$ is called {\em proper} when for each pair of vertices $x,y$ we have $|c(x)-c(y)|\ge w(x,y)$.
The number $s$ is called the {\em span} of $c$. The goal is to find a proper assignment of minimum span. 
Note that the special case when $w$ is $1$-bounded corresponds to the classical graph coloring problem.

In this paper we deal with exact algorithms for the channel assignment problem. 
As a generalization of graph coloring, the decision version of channel assignment is NP-complete. 
It follows that the existence of a polynomial-time algorithm is unlikely. 
As a consequence, researchers began to study exponential-time algorithms for the channel assignment problem. 
The asymptotic efficiency of these algorithms is measured in terms of $n=|V|$ and $\ell$, we assume that $\ell$ is a constant.
%and we assume that $\ell$ is a constant.
%PO CO ^ TO?
The first non-trivial algorithm was proposed by McDiarmid~\cite{mcdiarmid} and had running time of $O(n^2(2\ell+1)^n)$. It was then improved by Kral~\cite{kral} to $O(n(\ell+2)^n)$. 

Here we improve the running time further to $O^*((\ell+1)^n)$\footnote{By $O^*()$ we suppress polynomially bounded terms.}. We also show that the number of all proper assignments can be found in the same time bound.
Note that for $\ell=1$ the running time of our algorithm matches the time complexity of the currently fastest algorithm for graph coloring by Bj\"{o}rklund, Husfeldt and Koivisto~\cite{bhk}. 

Our improvement is achieved by applying the fast zeta transform in combination with the inclusion-exclusion principle.
The same ingredients were used also in a set partition problem in~\cite{bhk}, however in our algorithm the fast zeta transform plays a different role. In particular, although channel assignment resembles a kind of set partition it does not seem to be possible to solve it by a direct application of the algorithm from~\cite{bhk}.

Some special cases of the channel assignment problem received particular attention.
An important example is the $L(p,q)$-labeling of graphs, where given an undirected graph $G=(V,E)$ one has to find an assignment $c:V\rightarrow\mathbb{N}$ such that if vertices $u$ and $v$ are adjacent then $|c(u)-c(v)|\ge p$ and if vertices $u$ and $v$ are at distance 2 then $|c(u)-c(v)|\ge q$. The goal is to minimize $\max_{v\in V} c(v)$. Clearly, the algorithmic problem of finding an $L(p,q)$-labeling reduces in polynomial time to the $\max\{p,q\}$-bounded channel assignment and we get an $O^*((\max\{p,q\}+1)^n)$-time algorithm as an immediate corollary from our result. In particular, it gives an $O^*(3^n)$-time algorithm for the most researched subcase of $L(2,1)$-labeling. This improves over the algorithms by Havet et al.~\cite{havet} running in time $O(3.873^n)$ and a recent improvement of Junosza-Szaniawski and Rz\k{a}\.{z}ewski~\cite{pw} running in $O(3.562^n)$ time.

\section{Deciding}
\label{sec:decision}
In this section we consider the decision version of the problem, i.e.\ for a given $\ell$-bounded weight function $w$ and an integer $s\in\mathbb{N}$ we check whether there is a proper assignment of span at most $s$. Since the case $\ell=1$ can be solved in $O^*((\ell+1)^n)=O^*(2^n)$ time as described in~\cite{bhk}, here we assume $\ell \ge 2$.

An assignment $c:V\rightarrow\mathbb{N}$ of span $s$ can be seen as a tuple $(I_1,I_2,\ldots,I_s)$, where $I_j=c^{-1}(j)$ for every $j=1,\ldots,s$.
We will relax the notion of assignment in that we will work with tuples of vertex sets $(I_1,I_2,\ldots,I_k)$, where the $I_j$'s are not necessarily disjoint. We say that a tuple $(I_1,I_2,\ldots,I_k)$ is {\em proper}, when for every $i,j\in\{1,\ldots,k\}$ if $x\in I_i$ and $y\in I_j$ then $|i-j|\ge w(x,y)$.

In what follows, $U$ denotes the set of all proper tuples $(I_1,\ldots,I_s)$ such that for each $j=1,\ldots,s-\ell+1$, 
 the sets $I_j, I_{j+1}, \ldots, I_{j+\ell-1}$ are pairwise disjoint. 
%For a tuple $(I_1,\ldots,I_s)$ the {\em span} of the tuple is defined as the maximum index $j$ such that $I_j\neq\emptyset$.
A tuple with the $r$ last elements being empty sets is denoted as $(I_1,\ldots,I_{s-r},\emptyset^{r})$. 
% Note that this tuple is of span at most $s-r$.
%For convenience, for any $j<s$, a tuple $(I_1,\ldots,I_j,\emptyset,\ldots,\emptyset)\in U$ will be identified with the tuple $(I_1,\ldots,I_j)$, and in that sense we can write $(I_1,\ldots,I_j)$ for a tuple in $U$ of span at most $j$. 
For a subset $X\subset V$, we say that a tuple $(I_1,\ldots,I_j)$ {\em lies} in $X$ when for every $i=1,\ldots,j$, we have $I_i\subset X$.

For $v\in V$, define $U_v=\{(I_1,\ldots,I_s) \in U\ :\ v \in \bigcup_{j=1}^s I_j\}$.
Observe, that

\begin{proposition}
 $\displaystyle |\bigcap_{v\in V} U_v| > 0$ iff there is a proper assignment of span $s$.
\end{proposition}

By the inclusion-exclusion principle, if we denote $\overline{U_v}=U-U_v$ and $\bigcap_{v\in\emptyset}\overline{U_v}=U$, then 

\begin{equation}
\label{eq:i-e} 
|\bigcap_{v\in V} U_v| = \sum_{Y\subseteq V}(-1)^{|Y|}|\bigcap_{v\in Y}\overline{U_v}|.
\end{equation}

Our algorithm computes $|\bigcap_{v\in V} U_v|$ using the above formula. The rest of the section is devoted to computing $|\bigcap_{v\in Y}\overline{U_v}|$ for a given set $Y\subseteq V$. If we denote $X=V-Y$, then $\bigcap_{v\in Y}\overline{U_v}$ is just the set of tuples of $U$ that lie in $X$:

\begin{equation}
\label{eq:a} 
\bigcap_{v\in Y}\overline{U_v} = \{(I_1,\ldots,I_s)\in U\ :\ I_1,\ldots,I_s \subseteq X\}.
\end{equation}

Our plan now is to compute the value of $|\bigcap_{v\in Y}\overline{U_v}|$ using dynamic programming accelerated by the fast zeta transform.
More precisely, for every $i=\ell-1, \ldots, s$ and for every sequence $J_1,\ldots,J_{\ell-1}$ of pairwise disjoint subsets of $X$ our algorithm computes the value of 
\begin{equation}
\label{eq:T} 
T^X_i(J_1,\ldots,J_{\ell-1}) = |\{(I_1,\ldots,I_{i-(\ell-1)},J_1,\ldots,J_{\ell-1},\emptyset^{s-i})\in U\ :\ \bigcup_{j=1}^{i-(\ell-1)} I_j \subseteq X\}|,
\end{equation}
%I_1,\ldots,I_{i-(\ell-1)},J_1,\ldots,J_{\ell-1}
%that is, the number of tuples in $U$ of span at most $i$ that lie in $X$ and end with $J_1,\ldots,J_{\ell-1}$ followed by $s-i$ empty sets. Then, clearly,
that is, the number of tuples in $U$ that lie in $X$ and end with $J_1,\ldots,J_{\ell-1}$ followed by $s-i$ empty sets. Then, clearly,

\begin{equation}
\label{eq:res} 
|\bigcap_{v\in Y}\overline{U_v}| = \sum_{\substack{J_1,\ldots,J_{\ell-1}\subseteq X\\i\ne j\Rightarrow J_i \cap J_j = \emptyset}}T^X_s(J_1,\ldots,J_{\ell-1}).
\end{equation}

For every sequence of pairwise disjoint sets $J_1,\ldots,J_{\ell-1}\subseteq X$, we can initialize the value of $T^X_{\ell-1}(J_1,\ldots,J_{\ell-1})$ in polynomial time as follows\footnote{$[\alpha]$ is the Iverson's notation, i.e.\ $[\alpha]=1$ when $\alpha$ holds and $[\alpha]=0$ otherwise.}:

\begin{equation}
\label{eq:init} 
T^X_{\ell-1}(J_1,\ldots,J_{\ell-1}) = [(J_1,\ldots,J_{\ell-1})\text { is proper}].
%\begin{cases}
%                     1 & \text{when $(J_1,\ldots,J_{\ell-1})$ is proper}, \\ 
%                     0 & \text{otherwise.}
%                  \end{cases}
\end{equation}

Then the algorithm finds the values of $T^X_{j}$ for subsequent $j=\ell,\ldots,s$. This is realized using the following formula:
\begin{equation}
\label{eq:step} 
T^X_{i}(J_1,\ldots,J_{\ell-1}) = [(J_1,\ldots,J_{\ell-1})\text { is proper}]\cdot \sum_{J_0\subseteq X \cap \proper(J_1,\ldots,J_{\ell-1})}T^X_{i-1}(J_0,J_1,\ldots,J_{\ell-2}),
\end{equation}
where $\proper(J_1,\ldots,J_{\ell-1})$ is the set of all vertices $v\in V\setminus\bigcup_{j=1}^{\ell-1} J_j$ such that 
 for each $j=1,\ldots,\ell-1$ and $x\in J_j$ we have $j\ge w(v,x)$.

Using the formula~\eqref{eq:step} explicitly, one can compute all the values of  $T^X_{i}$ 
from the values of $T^X_{i-1}$ in $O^*((\ell+1)^{|X|})$ time, since there are $(\ell+1)^{|X|}$ tuples $(J_0,\ldots,J_{\ell-1})$ of disjoint subsets of $X$. Now we describe how to speed it up to $O^*(\ell^{|X|})$.

Let $S$ be a set and let $f:2^S\rightarrow \mathbb{Z}$ be a function on the lattice of all subsets of $S$.
The zeta transform is an operator which transforms $f$ to another function $(\zeta f):2^S\rightarrow \mathbb{Z}$ and it is defined as follows:
\[(\zeta f)(Q)=\sum_{R\subseteq Q} f(R).\]
A nice feature of the zeta transform is that given $f$ (i.e.\ when the value of $f(R)$ can be accessed in $O(1)$ time for any $R$) there is an algorithm (called fast zeta transform or Yates' algorithm, see~\cite{bhk,yates}) which computes $\zeta f$ (i.e.\ the values of $(\zeta f)(Q)$ for {\em all} subsets $Q\subseteq S$) using only $O(2^{|S|})$ arithmetic operations (additions).

Let us come back to our algorithm. In the faster version, for each $i=\ell,\ldots,s$, we iterate over all sequences of disjoint subsets $J_1,\ldots, J_{\ell-2}\subseteq X$. 
Then the values of $T^X_{i}(J_1,\ldots,J_{\ell-1})$ for all the $2^{|X|-\sum_{j=1}^{\ell-2}|J_j|}$ sets $J_{\ell-1}$ that are disjoint with $J_1,\ldots, J_{\ell-2}$ are computed in $O^*(2^{|X|-\sum_{j=1}^{\ell-2}|J_j|})$ time (that is in polynomial time per set!).
To this end, we use the function $f : 2^{X\setminus\bigcup_{j=1}^{\ell-2}J_j}\rightarrow \mathbb{Z}$, where
\[f(S) = T^X_{i-1}(S,J_1,\ldots,J_{\ell-2}).\]

We compute the function $(\zeta f)$ with the fast zeta transform using $O(2^{|X|-\sum_{j=1}^{\ell-2}|J_j|})$ additions. 
Now, observe that by~\eqref{eq:step}, for each $J_{\ell-1}\subseteq X$ disjoint with $J_1,\ldots, J_{\ell-2}$,  
\begin{equation*}
\label{eq:yates} 
T^X_{i}(J_1,\ldots,J_{\ell-1}) = [(J_1,\ldots,J_{\ell-1})\text { is proper}]\cdot(\zeta f)(X \cap  \proper(J_1,\ldots,J_{\ell-1})).
\end{equation*}

It follows that for each $i=\ell,\ldots,s$ the algorithm runs in time needed to perform the following number of additions:
\begin{equation}
\label{eq:time} 
O(\sum_{\substack{J_1,\ldots,J_{\ell-2}\subseteq X\\j\ne k\Rightarrow J_j\cap J_k=\emptyset}}2^{|X|-\sum_{j=1}^{\ell-2}|J_j|})=
O(\sum_{\substack{J_1,\ldots,J_{\ell-1}\subseteq X\\j\ne k\Rightarrow J_j\cap J_k=\emptyset}}1) = O(\ell^{|X|}).
\end{equation}

By~\eqref{eq:i-e} it follows that the whole decision algorithm runs in time needed to perform $O(n(\ell+1)^{n})$ additions.
The numbers being added are bounded by $|\bigcap_{v\in Y}\overline{U_v}|\le 2^{ns}\le 2^{n^2\ell}$, where the last inequality follows from
the fact that the minimum span is upper bounded by $(n-1)\ell +1$ (see e.g.~\cite{mcdiarmid}). Hence a single addition is performed in $O(n^2\ell)$ time.

\begin{corollary}
There is an algorithm which verifies whether the minimum span of an $\ell$-bounded instance of the channel assignment 
problem is bounded by $s$ which uses $O^*((\ell+1)^{n})$ time and $O^*(\ell^{n})$ space. 
\end{corollary}

%Let us note that the polynomial overhead caused by storing and adding large numbers can be easily avoided, at the cost of introducing a small error probability.
%Namely, for some number $t$, we pick a random integer $p\in [t,2t)$ and in our algorithm we store only numbers bounded by $p-1$, and we perform all arithmetic operations modulo $p$. If there is no assignment of span $s$, the algorithm always returns the correct answer, and otherwise it may be wrong with probability at most $\frac{1}{t}$. If $t$ is small enough to fit in $O(1)$ machine words, the algorithm runs in $O(n(\ell+1)^n)$ time and $O(\ell^{n})$ space. 

\section{Counting}

In this section we briefly describe how to modify the decision algorithm from Section~\ref{sec:decision} in order to make it {\em count} the number of proper assignments of span at most $s$.
We follow the approach of Bj\"{o}rklund et al.~\cite{bhk}.
The trick is to modify the definition of $U$. Namely, now every tuple $(I_1,\ldots,I_s)$ from $U_v$ {\em additionally} satisfies the following condition:
\begin{equation}
 \label{eq:count-u}
 \sum_{j=1}^s|I_j|=n.
\end{equation}
Observe, that then $|\bigcap_{v\in V} U_v|$ equals the number of proper assignments of span at most $s$.
Now, we add another dimension to the arrays $T^X_i$:
\begin{eqnarray*}
\label{eq:T2} 
T^X_{i,k}(J_1,\ldots,J_{\ell-1}) = |\{(I_1,\ldots,I_{i-(\ell-1)},J_1,\ldots,J_{\ell-1},\emptyset^{s-i})\in U\ :\ \bigcup_{j=1}^{i-(\ell-1)} I_j \subseteq X\\ \text{ and } \sum_{j=1}^{i-(\ell-1)}|I_j|+\sum_{j=1}^{\ell-1}|J_j|=k\}|.
\end{eqnarray*}
The dynamic programming algorithm from Section~\ref{sec:decision} can be easily modified to compute the values of $T^X_{i,k}(J_1,\ldots,J_{\ell-1})$ for all $i=\ell-1,\ldots,s$, $k=0,\ldots,n$ and all sequences of $\ell-1$ pairwise disjoint subsets of $X$. The details are left to the reader.

\begin{corollary}
For any $\ell$-bounded instance of the channel assignment problem the number of the proper assignments of span at most $s$ can be computed in $O^*((\ell+1)^{n})$ time and $O^*(\ell^{n})$ space. 
\end{corollary}

\section{Finding}

In order to find the assignment itself we can solve the extended version 
of the channel assignment problem, 
where we are additionally given a set of vertices $Z \subseteq V$
together with a function $c':Z \rightarrow \{1,\ldots,s\}$.
Then we are to check whether there exists a proper assignment $c:V\rightarrow \{1,\ldots,s\}$
satisfying $c |_Z = c'$.
It is not hard to modify the presented algorithm to solve the extended version of the problem 
in $O^*((\ell+1)^{n-|Z|})$ time.
The details are left to the reader.

Now using the extended version of the channel assignment problem we
can take any $v \in V \setminus Z$ and try each of the $s\le (n-1)\ell +1$ possible values of $c(v)$
one by one, each time using the algorithm for the extended channel assignment problem as a black box.
When the value for $v$ is fixed in a similar manner we assign the value for the other vertices of $V \setminus Z$.
Since $\sum_{i=1}^n(\ell+1)^{n-i}<(\ell+1)^n$, the algorithm for finding an assignment has a multiplicative overhead of $O(n\ell)$ over the running time of the decision version.

\section{Open problems}

In~\cite{traxler} Traxler has shown that for any constant $c$, the Constraint Satisfaction Problem (CSP) has no $O(c^n)$-time algorithm, assuming
the Exponential Time Hypothesis (ETH). More precisely, he shows that ETH implies that CSP requires $d^{\Omega(n)}$ time, where $d$ is the domain size.
On the other hand, graph coloring, which is a variant of CSP with unbounded domain, admits a $O^*(2^n)$-time algorithm. The channel assignment problem is a generalization of graph coloring and a special case of CSP. In that context, the central open problem in the complexity of the channel assignment problem is to find a $O^*(c^n)$-time algorithm for a constant $c$ independent of $\ell$ or to show that such the algorithm does not exist, assuming ETH (or other well-established complexity conjecture).

\bibliographystyle{abbrv}
\bibliography{channel}
\end{document}